\documentclass[prb,showpacs,twocolumn,superscriptaddress]{revtex4-1}

\usepackage{amsmath,amssymb,amsfonts,epstopdf,graphicx,enumerate,bm,float}

\pdfoutput=1

\newcommand{\beq}{\begin{eqnarray}}			\newcommand{\eeq}{\end{eqnarray}}
\newcommand{\bseq}{\begin{subequations}}		\newcommand{\eseq}{\end{subequations}}
\newcommand{\bem}{\begin{pmatrix}}			\newcommand{\eem}{\end{pmatrix}}
\newcommand{\ben}{\begin{enumerate}}			\newcommand{\een}{\end{enumerate}}
\newcommand{\bef}{\begin{figure}}			\newcommand{\eef}{\end{figure}}
\newcommand{\br}{\langle}					\newcommand{\ke}{\rangle}
\newcommand{\up}{\uparrow}		\newcommand{\down}{\downarrow}		\newcommand{\ra}{\rightarrow}
		\newcommand{\mb}{\mathbf}			\newcommand{\mbb}{\mathbb}
\newcommand{\nonu}{\nonumber\\}
\newcommand{\pa}{\partial}		\newcommand{\sig}{\sigma}			\newcommand{\Del}{\Delta}
\newcommand{\eps}{\epsilon}
	\def\f#1#2{\frac{#1}{#2}}

\begin{document}
\title{Tuning thermoelectric power factor by crystal-field and spin-orbit couplings in Kondo lattice materials}

\author{Seungmin Hong}
\affiliation{Department of Physics, University of Illinois at Urbana-Champaign, Urbana, IL 61801, USA}

\author{Pouyan Ghaemi}
\affiliation{Department of Physics, University of Illinois at Urbana-Champaign, Urbana, IL 61801, USA}
\affiliation{Department of Physics, University of California, Berkeley, CA 94720, USA}
\affiliation{Materials Sciences Division, Lawrence Berkeley National Laboratory, Berkeley, CA 94720, USA}

\author{Joel E. Moore}
\affiliation{Department of Physics, University of California, Berkeley, CA 94720, USA}
\affiliation{Materials Sciences Division, Lawrence Berkeley National Laboratory, Berkeley, CA 94720, USA}

\author{Philip W. Phillips}
\affiliation{Department of Physics, University of Illinois at Urbana-Champaign, Urbana, IL 61801, USA}

\begin{abstract}
We study thermoelectric transport at low temperatures in correlated Kondo insulators, motivated by the recent observation of a high
thermoelectric figure of merit(ZT) in  $FeSb_2$ at $T \sim 10
K$\cite{bentien07}.  Even at room temperature, correlations have the
potential to lead to high ZT, as in $YbAl_3$, one of the most widely
used thermoelectric metals.  At low temperature correlation effects
are especially worthy of study because fixed band structures are
unlikely to give rise to the very small energy gaps $E_g \sim 5 kT$
necessary for a weakly correlated material to function efficiently at
low temperature.  We explore the possibility of improving the
thermoelectric properties of correlated Kondo insulators through
tuning of crystal field and spin-orbit coupling and present a
framework to design more efficient low-temperature thermoelectrics
based on our results.
\end{abstract}

\date{}
\maketitle
\section{introduction}

Thermoelectrics support a voltage drop in response to a modest
temperature gradient.  Since a temperature gradient affects the
electrons and the lattice degrees of freedom, optimizing
thermoelectrics involves not only the thermopower or Seebeck
coefficient ($S$), but also the electrical ($\sigma$) and thermal $(\kappa)$ conductivities.  The holy grail of thermoelectrics is to achieve a figure of merit
\beq
ZT=\left(\frac{S^2\sigma}{\kappa}\right)T
\eeq
that exceeds unity at room temperature. This tall order remains a grand challenge problem\cite{pouyan10,dreselhause,wood,chemrev}.
Two promising recent directions have focused on either decreasing the thermal conductivity as in the case of nanocrystalline arrays of Bi$_x$Sb$_{2-x}$Te$_3$ in which a $ZT$ of 1.4 was achieved\cite{dreselhause} at $T=373K$ or maximizing the numerator of $Z$ through strong electron correlations.  An example of the latter is the report\cite{bentien07} that FeSb$_2$ achieves a colossal thermopower of $45000\mu V/K$ at $10K$ resulting in the largest power factor, $S^2\sigma$ witnessed to date.  In this paper, we follow-up on the role strong correlations play in maximizing the power factor by focusing on Kondo insulators.  We show explicitly that multi-orbital physics in Kondo insulators lies at the heart of the problem of maximizing the power factor. 

Because the thermopower is related to the entropy per carrier, particle-hole asymmetry and large density of states at the chemical potential are central to the optimization of $Z$.  In this regard, the Anderson model of a single impurity in a metal\cite{anderson}, which is among the few solvable strongly correlated systems solvable exactly, presents a density of states with demanding features for efficient thermoelectric transport.  For a single $SU(2)$ spin on a localized impurity, the density of states appears as a single infinite symmetric peak at the chemical potential leading to a divergent density of states but vanishing Seebeck coefficient by virtue of the particle-hole symmetry. Increasing the degeneracy of the localized orbital and the metallic band to $SU(N)$ ($N>2$) softens the peak in the density of states  and at the same time moves above the chemical potential leading to an asymmetric density of states and as a result a larger Seebeck coefficient\cite{hewson}.

It makes sense then to consider systems in which such physics is naturally present, for example Kondo insulators in which a regular lattice of Anderson impurities is hybridized with multiple bands of itinerant electrons. The electrons in the local orbitals are poorly screened and the strong Coloumb repulsion prohibits them from being multiply occupied. Contrary to the single impurity, the periodic Anderson model is not exactly solvable but multiple mean-field type methods were used\cite{read84,millis87,read87} to understand many of their features. Motivated by the single impurity model, we examine the effect of degeneracy of the local impurities and the conduction band on the thermoelectric properties of Kondo insulators. In addition to directly studying the degeneracy of the local and conduction bands, we study the effect lifting the degeneracy by a crystal-field (which mainly effects the local orbitals) and spin-orbit coupling (which mainly effects the conduction band) have on thermoelectric efficiency. In this way, we can continuously lift the level of degeneracy. Interestingly, we observe that there is an optimum value of the crystal field and spin orbit coupling. As was shown in a previous study \cite{kotliar}, the presence of multiple orbitals close to the chemical potential is a common feature of Kondo insulators. Our results indeed present a possible route for using strong correlations to enhance the thermoelectric performance through controlling the orbital degeneracy of local and itinerant bands.  

\section{Model and methodology}

Heavy fermion materials typically contain rare earth or actinide ions
forming a lattice of localized magnetic moments\cite{piers}. The
strong Coulomb repulsion of electrons localized in $f$ or $d$ orbitals
leads to formation of these local moments\cite{anderson} which then
hybridize with the itinerant electron bands and form the heavy
electron bands.  If the chemical potential is in the heavy electron
bands, a heavy fermion metal is formed. The volume of the Fermi
surface in this correlated state corresponds to a sum of the number of
itinerant and localized electrons. If the chemical potential is in the
hybridization gap, the heavy electron band will be fully occupied and
a Kondo insulator obtains\cite{kins1,kins2}. Notice that such an
insulating state is fundamentally different from a non-interacting
insulator.  For example, in order to reach a filled valance band, we need to add the number of localized and itinerant electrons which is solely developed as a result of strong interactions.

 The underlying microscopic model of this correlated system is:
\beq\label{mh}
H &=& \sum_{\mb kl\sig\sig'}\varepsilon_{\sig\sig'}(\mb k)c_{\mb kl\sig}^\dag c_{\mb kl\sig'}
	+ \sum_{\mb kl\sig} \eps_{f_l} d_{\mb kl\sig}^\dag d_{\mb kl\sig} +
	\nonu
  & &
 \sum_{i\mb kl\sig}\left(
		V_{\mb kl\sig}e^{i\mb k\cdot\mb r_i}c_{\mb kl\sig}^\dag d_{il\sig}+\text{h.c.}\right)
\nonu
 & &  	
+ \f U2\left(\sum_{il,\sig\neq\sig'} n_{il\sig}^d n_{il\sig'}^d+\sum_{i,l\neq l',\sig\sig'} n_{il\sig}^d n_{il'\sig'}^d\right)
\eeq
where $c_{\mb kl\sig}^\dag$($d_{\mb kl\sig}^\dag$) is the creation of
a conduction(local) electron with momentum $\mb k$, orbital $l$, and
spin $\sig=(\up, \down)$, and $n_{il\sig}^d=d_{il\sig}^\dag
d_{il\sig}$ is the number operator of a local $d$ orbital at site $\mb
r_i$. The dispersion of the $c$-electron $\varepsilon_{\sig\sig'}(\mb
k) = \eps_{\mb k}\delta_{\sig\sig'}+\bm\Gamma_{\mb
  k}\cdot\bm\sig_{\sig\sig'}$ includes spin coupling. The
non-dispersive energy of local states ($\eps_{f_l}$) depends on the
orbital index $l$. The pseudovector $\bm\Gamma_{\mb k}$ represents the
amplitude of the spin-orbit (SO) coupling\cite{SamokhinK09,Isaev} and
its form depends on the crystal symmetry of the underlying lattice
(see appendix \ref{sp}). Typically, the hybridization matrix element,
$V_{\mb kl\sig}$ encodes the complex orbital structures of local
states which can have novel effects on the properties of the strongly
correlated heavy fermion phase\cite{pouyan08}, but as in other studies, we consider $V_{\mb kl\sig}$ to be independent of $(\mb k,\sig)$ to make the calculation more tractable. 

Using the model Hamiltonian, Eq.~\eqref{mh}, we can capture the effect
of the degeneracy of both localized and itinerant bands, as well as
the effect of crystal field and spin orbit coupling in breaking the
degeneracy
 of these bands.
As a result of weak screening of electrons in $f$ and $d$ orbitals,
the associated on-site repulsive potential $U$ is much larger than the
hopping energies of the itinerant electrons.  To treat the large on-site repulsion term, we use the $U(1)$ slave-boson mean-field theory\cite{millis87,mao99}. In this treatment, the creation operator of a local electron $d_{il\sig}^\dagger=f^\dag_{il\sig}b_{i}$ is partitioned into a neutral fermion $f^\dag_{il\sig}$, and a charged boson $b_i$ that accounts for annihilation of an empty state. Since the local Hilbert space is restricted to either an empty or a singly occupied state, the additional local constraint,
\beq
\tilde Q_i=b_i^\dag b_i+\sum_{l\sig}f_{il\sig}^\dag f_{il\sig}=1
\eeq
should be enforced at every site $\mb r_i$. The Hamiltonian in terms of these slave particles then becomes

\begin{equation}\begin{split}\label{hamiltonian} 
H =& \sum_{\mb kl\sig\sig'}\varepsilon_{\sig\sig'}(\mb k)c_{\mb kl\sig}^\dag c_{\mb kl\sig'}
	+ \sum_{\mb kl\sig} \eps_{f_l} f_{\mb kl\sig}^\dag f_{\mb kl\sig} \\
	& + \sum_{i\mb kl\sig}\left(
		V_l^*e^{-i\mb k\cdot\mb r_i} f^\dag_{il\sig}b_i c_{\mb kl\sig}+\text{h.c.}\right)
  	+ \sum_i \lambda_i(\tilde Q_i-1)
\end{split}\end{equation}
where $\lambda_i$ is a Lagrange multiplier to maintain the local
constraint. In the above Hamiltonian, the effect of the crystal field
is to break the degeneracy of the local orbital states $ \eps_{f_l}$
whereas the spin-orbit coupling 
breaks the spin degeneracy of the conduction band. As a result, by tuning the crystal field and spin-orbit coupling, we can change the degeneracy of the local and conduction orbitals in a continuous manner.  Consequently, we have a tunable knob to gain the optimum thermoelectric performance.



The mean field approximation to the model Hamiltonian can be obtained by taking the coherent expectation $b=\br b_i\ke = \br b_i^\dag \ke$ and $\lambda = \br\lambda_i\ke$. This replacement effectively renormalizes the mixing matrix element $V_l\ra bV_l$, and the local energy $\eps_{f_l}\ra \eps_{f_l}+\lambda$ and leads to the quadratic Hamiltonian
\begin{equation}\begin{split}
H_{\text{MF}} =& \sum_{\mb klh}(\eps_{\mb k}+h|\bm\Gamma_{\mb k}|)c_{\mb klh}^\dag c_{\mb klh} 
	+ \sum_{\mb kl\sig} (\eps_{f_l}+\lambda) f_{\mb klh}^\dag f_{\mb klh} \\
	+ & \sum_{\mb klh}\left(
		bV_l^* f^\dag_{\mb klh} c_{\mb klh}+\text{h.c.}\right)
  	+ \lambda\sum_i (b^2-1).
\end{split}\end{equation}
Instead of working in the spin basis, we will use a helical basis that diagonalizes the single-electron dispersion $\varepsilon_{\sig\sig'}(\mb k)\ra \left[U_{\mb k}^\dag\varepsilon(\mb k)U_{\mb k}\right]_{hh'}=\left(\eps_{\mb k}+h|\bm\Gamma_{\mb k}|\right)\delta_{hh'}$ with $h,h'=\pm 1$. Then $c_{\mb klh}$($f_{\mb klh}$) is accordingly rotated by the unitary matrix $U_{\mb k}$ from the spin basis, $c_{\mb kl\sig}$($f_{\mb kl\sig}$). By performing the Bogoliubov transformation,
\begin{eqnarray}
a_{\mb klh+}&=& \alpha_{\mb klh}c_{\mb klh}+\beta_{\mb klh}f_{\mb klh}\\ 
a_{\mb klh-} &=&-\beta_{\mb klh}c_{\mb klh}+\alpha_{\mb klh}f_{\mb klh},
\end{eqnarray}
we obtain the diagonal mean-field Hamiltonian,
\beq
H_{\text{MF}} = \sum_{\mb klh,\pm} E_{\mb klh}^{\pm} a_{\mb klh\pm}^\dag a_{\mb klh\pm} + \lambda\sum_i (b^2-1),
\eeq
where the dispersion is given by 
\beq
E_{\mb klh}^{\pm} 
	&=&	\f12(\eps_{\mb k}+h|\bm\Gamma_{\mb k}|+\eps_{f_l}+\lambda\pm W_{\mb klh}),\\
W_{\mb klh}
	&=&\sqrt{(\eps_{\mb k}+h|\bm\Gamma_{\mb k}|-\eps_{f_l}-\lambda)^2+4b^2V_l^2}.
\eeq
The Bogoliubov parameters are
\beq
\bem\alpha_{\mb klh}^2\\ \beta_{\mb klh}^2\eem 
	= \f12\left[1 \pm \f{(\eps_{\mb k}+h|\bm\Gamma_{\mb k}|)-(\eps_{f_l}+\lambda)}{W_{\mb klh}}\right].
\eeq

Minimization of the free energy with respect to the mean field parameters $b$ and $\lambda$, and the total chemical potential $\mu$ leads to two coupled equations
\beq
1 &=& b^2+\sum_{\mb klh}\alpha_{\mb klh}^2 n_F(E_{\mb klh+})+\beta_{\mb klh}^2 n_F(E_{\mb klh-}),\\
\lambda &=& \sum_{\mb klh} \f{V_l^2}{W_{\mb klh}}\big[n_F(E_{\mb klh-})-n_F(E_{\mb klh+})\big],\\
n_{\text{tot}} &=& \sum_{\mb klh}\big[n_F(E_{\mb klh-})+n_F(E_{\mb klh+})\big],
\eeq
where, the total density of electrons is fixed to be $n_{\text{tot}}=2 l_{\text{max}}$ for $l=1, 2, \cdots, l_{\text{max}}$. The transport of this non-interacting mean-field Hamiltonian is now tractable.

To compute the transport properties, we use the relaxation-time approximation to the Boltzmann equation\cite{ash}. Under this scheme, the electrical resistivity, $\bm\rho=\bm\sig^{-1}$, and the thermopower tensors, $\mb S$, are given by
\beq
\bm\rho = \mb L_0^{-1}\ ,\qquad \mb S=-\f{k_B}{|e|}\mb L_0^{-1}\mb L_1,
\eeq
where the tensors $\mb L_m$ are 
\beq\begin{split}
(\mb L_m)_{ab} = & -\f{e^2}{V_{\text{olume}}}\sum_{\mb k l h\pm}
	\f{\pa n_F(E_{\mb klh\pm})}{\pa E_{\mb klh\pm}}
	\\ & \times \tau_{\mb klh\pm}(\mb v_{\mb klh\pm})_a(\mb v_{\mb klh\pm})_b
	\left(\f{E_{\mb klh\pm}-\mu}{k_BT}\right)^m,
\end{split}\eeq
explicitly.  Here we set $\mb v_{\mb klh\pm}=\f1\hbar\bm\nabla E_{\mb klh\pm}$. Considering that the electrons are scattered by $N_{\text{imp}}$ impurities with an interaction strength of $V_{\text{imp}}$ ({\it i.e.,} $H_{\text{sctt}}\sim V_{\text{imp}}c_{\mb k'l\sig}^\dag c_{\mb kl\sig}$), the relaxation time $\tau_{\mb klh\pm}$ for each state is given by
\beq
\f1{\tau_{\mb klh\pm}} = \f{2\pi}\hbar \f{N_{\text{imp}}}{N_{\text{site}}}|V_{\text{imp}}|^2
	\left[\f{\pa E_{\mb klh\pm}}{\pa (\eps_{\mb k}+h|\bm\Gamma_{\mb k}|)}\right]^2
	\rho_{lh}(E_{\mb klh\pm}),\nonumber \\
\eeq
with $\rho_{lh}(E_{\mb klh\pm})$ the density of the states of the Bogoliubov quasiparticles.

\section{results}

We now present our results on the dependence of the transport properties on the orbital degeneracies of both localized and itinerant electron bands which form correlated Kondo insulators.
In the first two subsections, we consider double degeneracy of
conduction and localized bands. This model is indeed consistent with
the models previously proposed for Kondo insulators\cite{kontani}. The
crystal field will then split the degeneracy of the two $f$ levels in
\ref{hamiltonian} and spin-orbit coupling breaks the degeneracy of the
conduction band states with differing helicity\cite{Isaev}. In these
two 
sections, we change the size of the degeneracy-breaking gap continuously. Using the
relaxation-time approximation, we can then calculate the transport
properties of a Kondo insulator. For most of the materials, the
dominant contribution to the thermal conductivity comes from 
lattice vibrations; as a consequence, the electronic contribution to the thermoelectric performance is measured through the power factor $Z_{\text{PF}}=\sigma S^2$ where $\sigma$ is the electrical conductivity and $S$ is the Seebeck coefficient. 
In order to confirm that the enhancement of the thermoelectric
efficiency can properly be attributed to strong correlations, we consider two different band structures of itinerant electrons in \ref{pr} and \ref {tb} and we see that similar features emerge. 

Finally,  in \ref{mr} we present the effect of multiple orbital degeneracy. Contrary to the treatment in \ref{pr} and \ref{tb} where the double degeneracy is continuously lifted by crystal field and spin-orbit couplings, in section \ref{mr} we discretely change the number of degenerate conduction and localized bands. We show that indeed there is also an optimum degeneracy associated with the maximum power factor.

\subsection{Nearly free electron itinerant bands}\label{pr}

 We first focus on the effect of crystal field and spin-orbit coupling
 on the power factor within the context of a parabolic band for the
 itinerant electrons $\eps_{\mb k}=\eps_0 + W(k/k_{BZ})^2$ with
 $W=2$eV, taken from ref.~\cite{Carlos96}. In principle, the
 SO coupling should be expressed as a periodic function under the
 crystal environment, but we model it to be isotropic as well,
 $\bm\Gamma_{\mb k}=\gamma_{\text{so}} (\mb k/k_{BZ})$
 ($\gamma_{\text{so}}\leq 0.2\, eV$). In the following, we carry out
 the numerical calculation based on this isotropic band dispersion
 with $l_{\text{max}}=2$. Here, we choose $\eps_{f_1}=1.0606$eV,
 $V_1=0.2236$eV and $V_2=1.05V_1=0.2348$eV. The parameter
 $\f{N_{\text{imp}}}{N_{\text{site}}}|V_{\text{imp}}|^2=0.045$eV$^2$\cite{Carlos96}. The
 other control parameters are temperature($T\leq 100K$), crystal
 electric field (CEF)
 splitting($\Del_{\text{CEF}}=\eps_{f_2}-\eps_{f_1}\leq 15$meV), and
 the SO coupling($\gamma_{\text{so}}\leq0.15\,eV$). Although we do not 
include the supporting data here, we found that our conclusions are insensitive to the strength of $V_2$ as long as $0.5\lesssim V_2/V_1\lesssim 2.0$.

\begin{figure}[!t]
\begin{center}
\includegraphics[width=1.0\columnwidth]{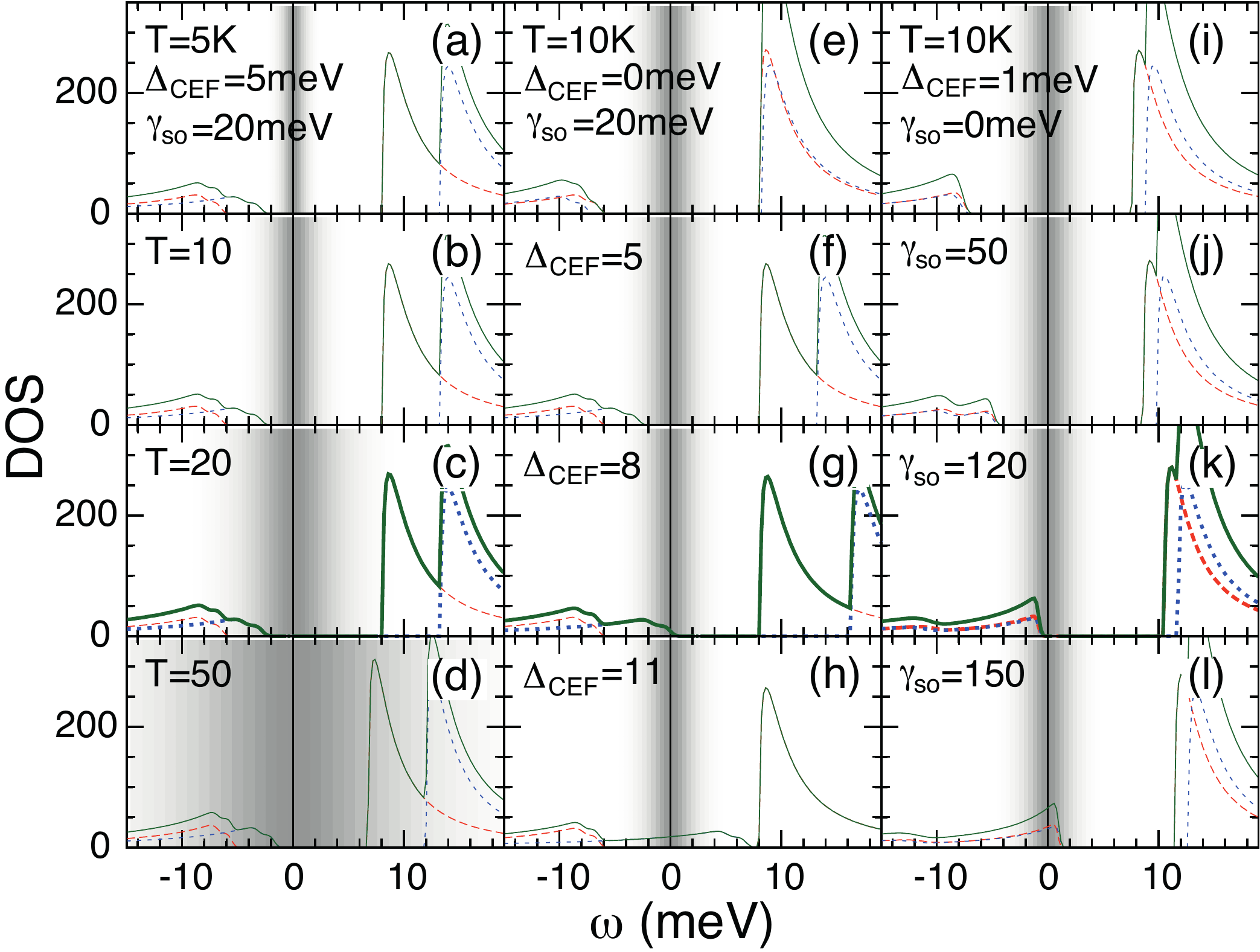}
\caption{Density of states (DOS) of the parabolic model for different
  control parameters. The red-dashed lines are for the orbital $l=1$,
  the dotted blue lines  $l=2$, and the green lines are for the total
  DOS. The central gray area displays the thermal window($\sim k_BT$)
  for each temperature. Figures (a)-(d) compare the DOS for
  different temperatures. Figures (e)-(h) correspond to different
  magnitudes of the crystal electric field($\Del_{\text{CEF}}$)  which
  breaks the degeneracy of the two $f$ orbitals. Figures (i)-(l)
  correspond to different magnitudes of the spin-orbit interaction($\gamma_{\text{SO}}$) that breaks the degeneracy of the conduction bands with different helicity.}\label{Fig1}
\end{center}
\end{figure}

\begin{figure}[!t]
\begin{center}
\includegraphics[width=0.99\columnwidth]{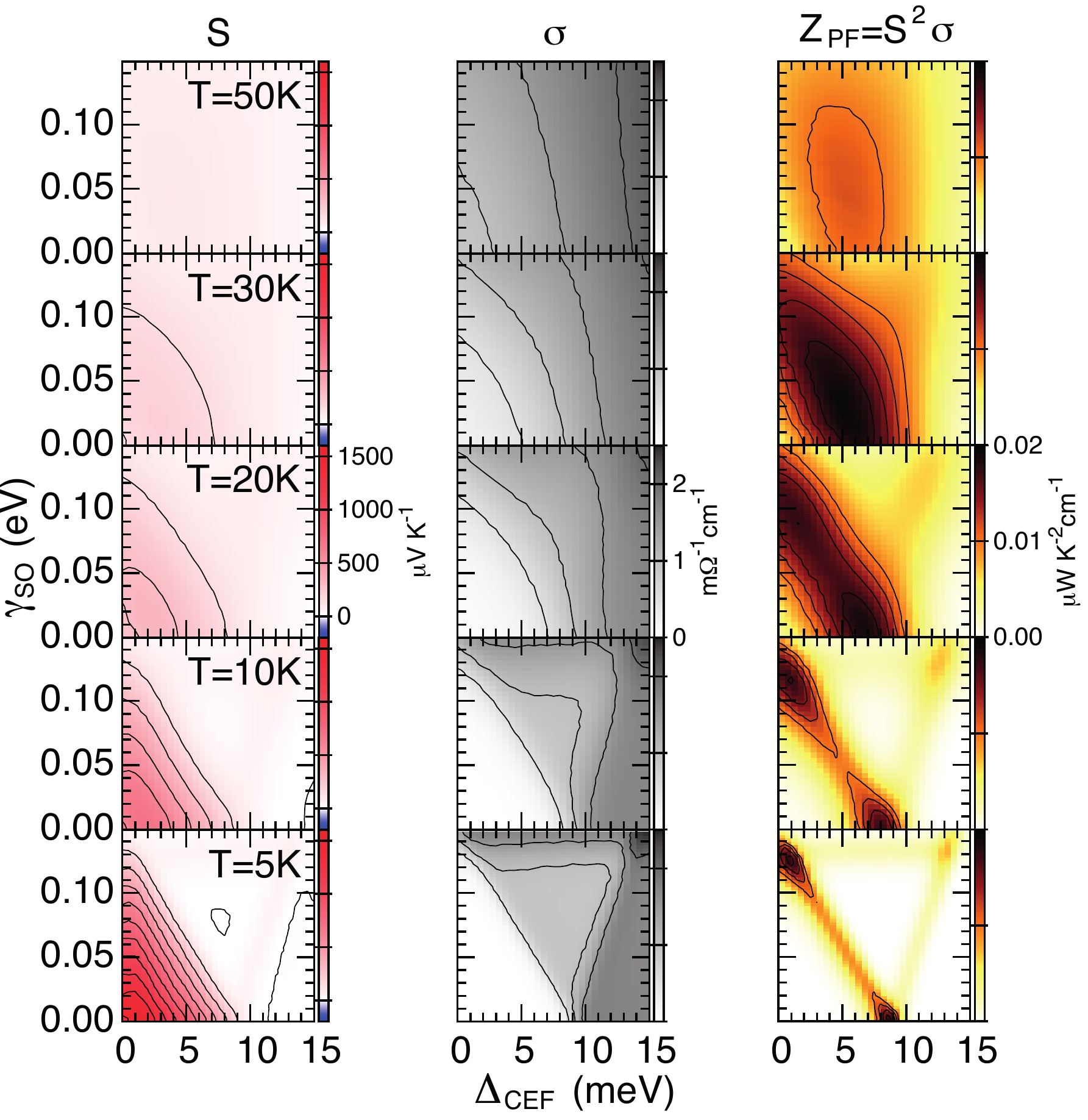}
\caption{Transport coefficients for different temperatures as a function of crystal field splitting and spin-orbit interaction: thermopower $S$ (left column), conductivity $\sig$ (middle), and power factor $Z_{\text{PF}}$ (right). From the top to the bottom, the temperature varies from 50 to 5K, and the solid lines are equally spaced constant contours.}\label{Fig2}
\end{center}
\end{figure}

Figure $\ref{Fig1}$ shows the density of states (DOS) for different
control parameters. From Figs.~$\ref{Fig1}$(a) to (d), we notice
that the temperature only controls the number of thermally activated
charge carriers, while it does not significantly change the DOS
compared to the other parameters. When the degeneracy of the two local
orbitals is broken by the crystal electric field, one of the
hybridized bands moves closer to the chemical potential. Consequently,
the system is driven from an insulator to a conductor
[Figs.~\ref{Fig1}(e)-(h)], at which point the power factor is
significantly enhanced (See Fig.~\ref{Fig2}). Likewise, the spin-orbit
interaction breaks the degeneracy of the two helical modes, which
turns an insulator into a metallic state [Figs.~\ref{Fig1}(i)-(l)]. We
point out that the metallic state is characterized either by a local orbital, $l=2$ [Fig.~\ref{Fig1}(h)] or by a helicity, $h=+$ [Fig.~\ref{Fig1}(l)], since only the bands with corresponding quantum numbers are conducting. 

In order to further examine the effect of the control parameters, we
first calculate the transport coefficients as a function of the CEF
and SO. In Fig.~\ref{Fig2}, we show the results of a calculation of 
the thermopower ($S$), the electrical conductivity ($\sigma$), and the power factor
($Z_{\text{PF}}$). As can be seen from the right column, the power
factor is enhanced either by finding the optimal CEF or by adjusting
the SOI. Since both CEF and SOI shift some of the lower energy bands
toward the chemical potential [Figs.~\ref{Fig1}(e)-(l)], the number
of lower energy bands relevant for thermal transport is controlled by
CEF and SOI simultaneously. For the temperature range $T\lesssim 20K$,
where the thermal windows are sufficiently narrow, CEF and SOI
compete;  hence there are two distinctive optimal regimes. For a
sufficiently wide thermal window, attainable at intermediate
temperatures, $T\sim 30K$, CEF and SOI are working cooperatively to
form a single optimal region. For $T>30K$, the enhancement in
$Z_{\text{PF}}$ is not as drastic as at low
temperature. $Z_{\text{PF}}$ is maximized in the vicinity of the
insulator-metal transition (see the conductivity $\sig$ at $T=5-20K$),
resulting from a competition between $S$ and $\sig$. For instance, at
$T=5K$, the thermopower $S$ decreases with SOI and CEF, while the
system acquires a finite conductivity. Note that the metallic state here has one dominant helical state over the other. 

\begin{figure}[!t]
\begin{center}
\includegraphics[width=1.0\columnwidth]{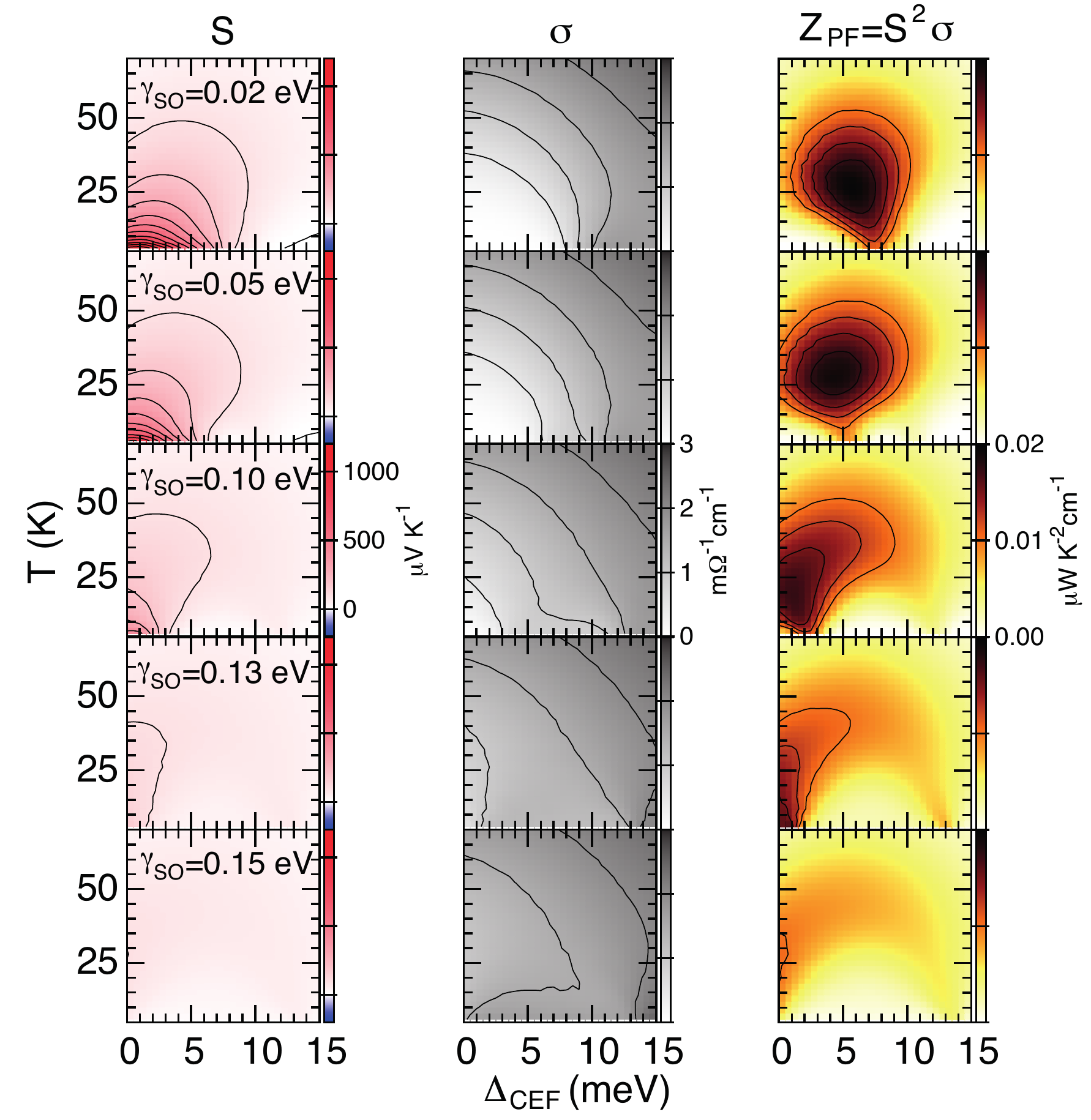}
\caption{Transport coefficients for each SOI as a function of CEF and temperature: thermopower $S$ (left column), conductivity $\sig$ (middle), and power factor $Z_{\text{PF}}$ (right). The range of SOI is $0.02-0.15$ eV from the top to bottom panels.}\label{Fig3}
\end{center}
\end{figure}

In Fig.~\ref{Fig3}, we repeat the calculation of the transport
coefficients for a fixed SOI as a function of CEF and
temperature. Consistent with Fig.~\ref{Fig2} is that the optimal point
for the power factor is located in the vicinity where the
insulator-metal transition occurs. For instance, when
$\gamma_{\text{so}}\lesssim 0.1$eV, there is the optimal CEF and
temperature for the power factor, at which point the electric
conductivity acquires a noticeable finite value. From the left
column, one finds that the thermopower generally decreases with
increasing temperature as a widened thermal widow implies the reduction of the asymmetry in the DOS within the thermal region [Figs.\ \ref{Fig1}(a)-(d)]. This obtains because as the temperature increases, more of the bands (lower and upper) are involved in the thermal transport. In other words, the asymmetry of the DOS within the thermally active region is relieved. Beyond a certain of the threshold SOI, $\gamma_{\text{so}}\ge 0.13$eV, there is no phase transition (at mean field level); hence optimization cannot be realized.

\subsection{Tight binding itinerant electron bands}\label{tb}
Next, we consider the 3-dimensional tight binding case. We see that as
in the case of a quadratic band, tuning the crystal-field and
spin-orbit coupling can optimize the thermoelectric performance. This
result indicates that the effect of orbital degeneracy in controlling
the thermoelectric performance is not that sensitive to the details
of the band structure.  Here, we choose the hopping parameter $t_{\text{hop}}=0.2167$eV (band width $W=2.6$eV), and we located the local energy $\eps_{f_1}=-0.8t_{\text{hop}}$. The hybridization strength $V_1=t_{\text{hop}}$ and $V_2=1.01t_{\text{hop}}$. In the SOI, we take the next-nearest neighbor hopping parameter $g_2=0.3$.

\begin{figure}[!b]
\begin{center}
\includegraphics[width=1.\columnwidth]{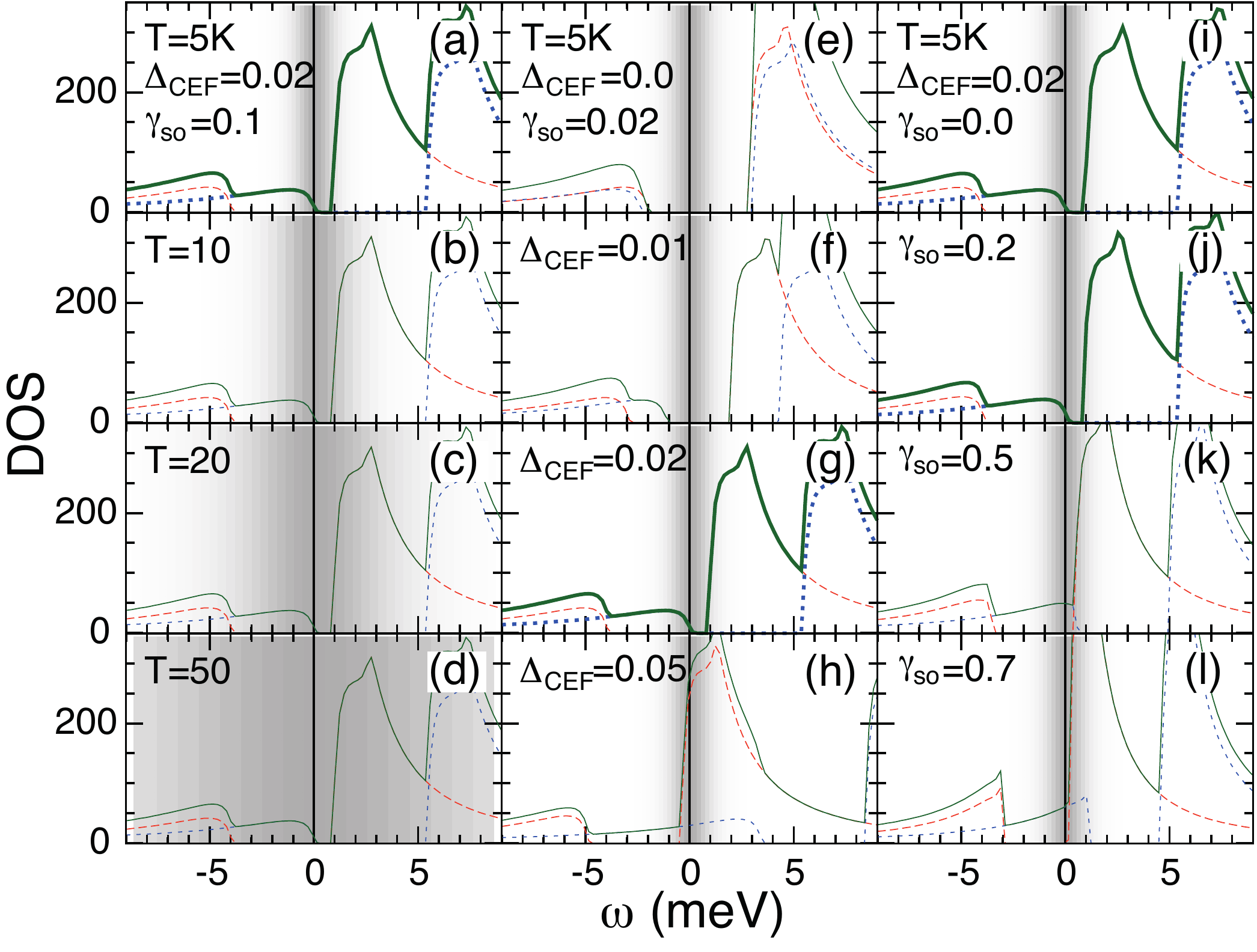}
\caption{Density of states(DOS) of the 3D tight-binding model for
  different control parameters. Both $\Del_{\text{CEF}}$ and
  $\gamma_{\text{so}}$ are in units of the hoping amplitude,
  $t_{\text{hop}}=0.216$eV. The red-dashed lines are for the
 orbital $l=1$, the dotted blue lines are for $l=2$, and the green lines are for the total DOS. The central gray area indicates the thermally active region for each temperature. Figs.~(a)-(d) compare the DOS for different temperature, (e)-(h) for the crystal electric field($\Del_{\text{CEF}}$), and (i)-(l) for the spin-orbit interaction($\gamma_{\text{SO}}$).}\label{Fig4}
\end{center}
\end{figure}

\begin{figure}[!t]
\begin{center}
\includegraphics[width=1.\columnwidth]{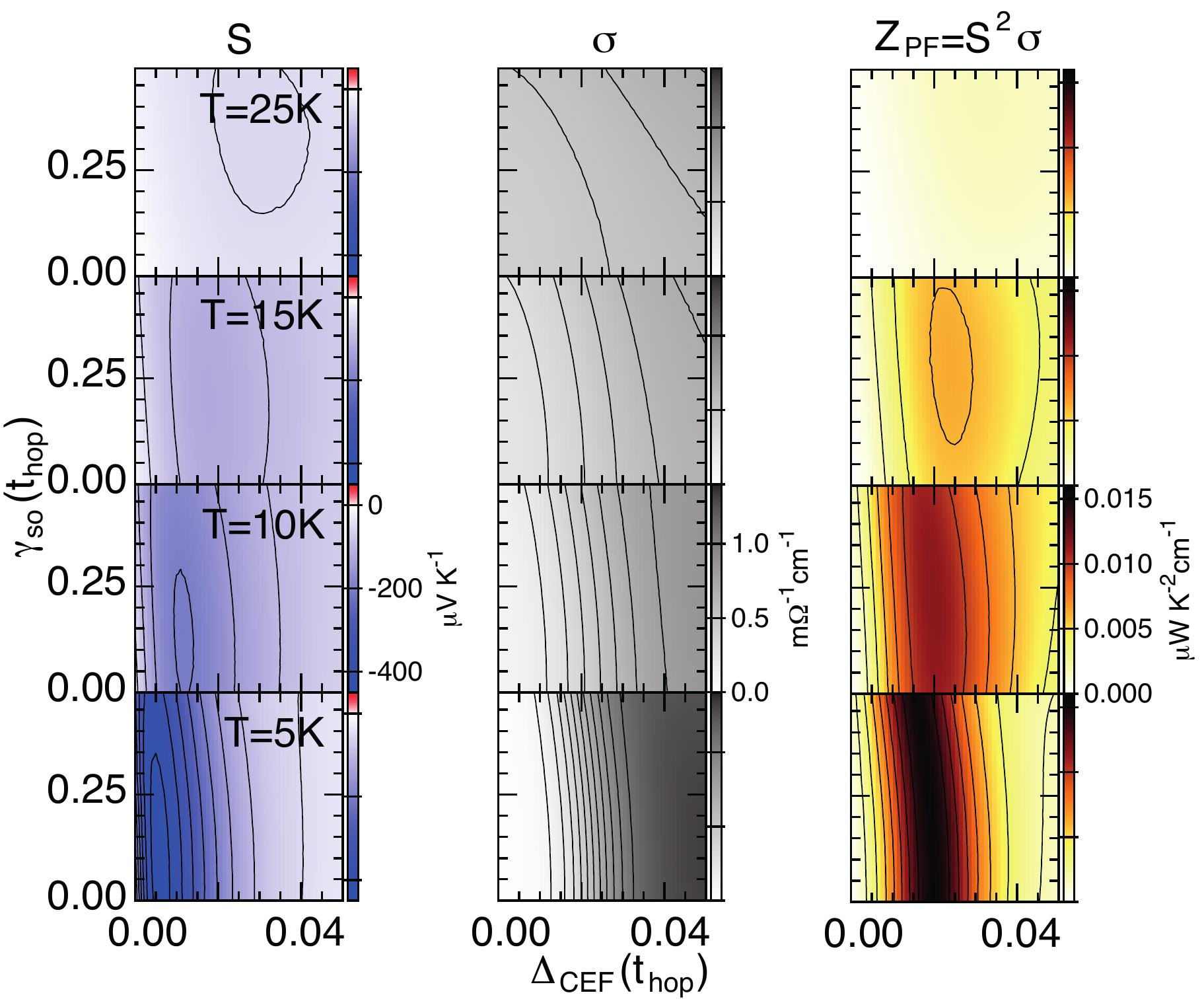}
\caption{Transport coefficients for each temperature: thermopower $S$ (left column), conductivity $\sig$ (middle), and power factor $Z_{\text{PF}}$ (right). From the top to the bottom, the temperature is fixed to 25, 15, 10, and 5K, respectively. }\label{Fig5}
\end{center}
\end{figure}

As in the simplified parabolic model, the role of CEF and SOI is not
different; both efficiently control the system to drive it from an
insulator to a conductor as seen from Fig.~\ref{Fig4}. Compared to the
corresponding panels in Figs.~\ref{Fig1}, however,
Figs.~\ref{Fig4}(e)-(h) show that the CEF also pushes one of the
upper bands toward the chemical potential, hence reducing the gap size
significantly. In fact, the parabolic model is rather exceptional
since the bottom of the upper bands corresponds to the point $\mb
k=\mb 0$, which is not usual for typical 3D tight-binding
models. Figs.~\ref{Fig4}(i)-(l) display the evolution of DOS with
the increase of the SOI. Even though the degeneracy of the
two helical modes are broken with a finite SOI, it cannot be seen
clearly as was in the linearized SOI case
[Figs.~\ref{Fig1}(i)-(l))]. The reason is that $|\bm\Gamma_{\mb k}|$
decreases as $\mb k$ approaches the boundary of the Brillouin zone due
to the periodic form of the SOI, while it does not for the linearized
SOI.  Unlike the CEF which affected drastically  only one of the
orbitals, the effect of the SOI is quite different. Up to
$\gamma_{\text{so}}=0.2t_{\text{hop}}\simeq 40$meV, the changes in the
DOS are not significant. For $\gamma_{\text{so}}\gtrsim 0.2t_{\text{hop}}$, the system undergoes a phase transition to a (helically polarized) metal, beyond which point $Z_{\text{PF}}$ is reduced (See Fig.~\ref{Fig5}). 

\begin{figure}[!t]
\begin{center}
\includegraphics[width=1.\columnwidth]{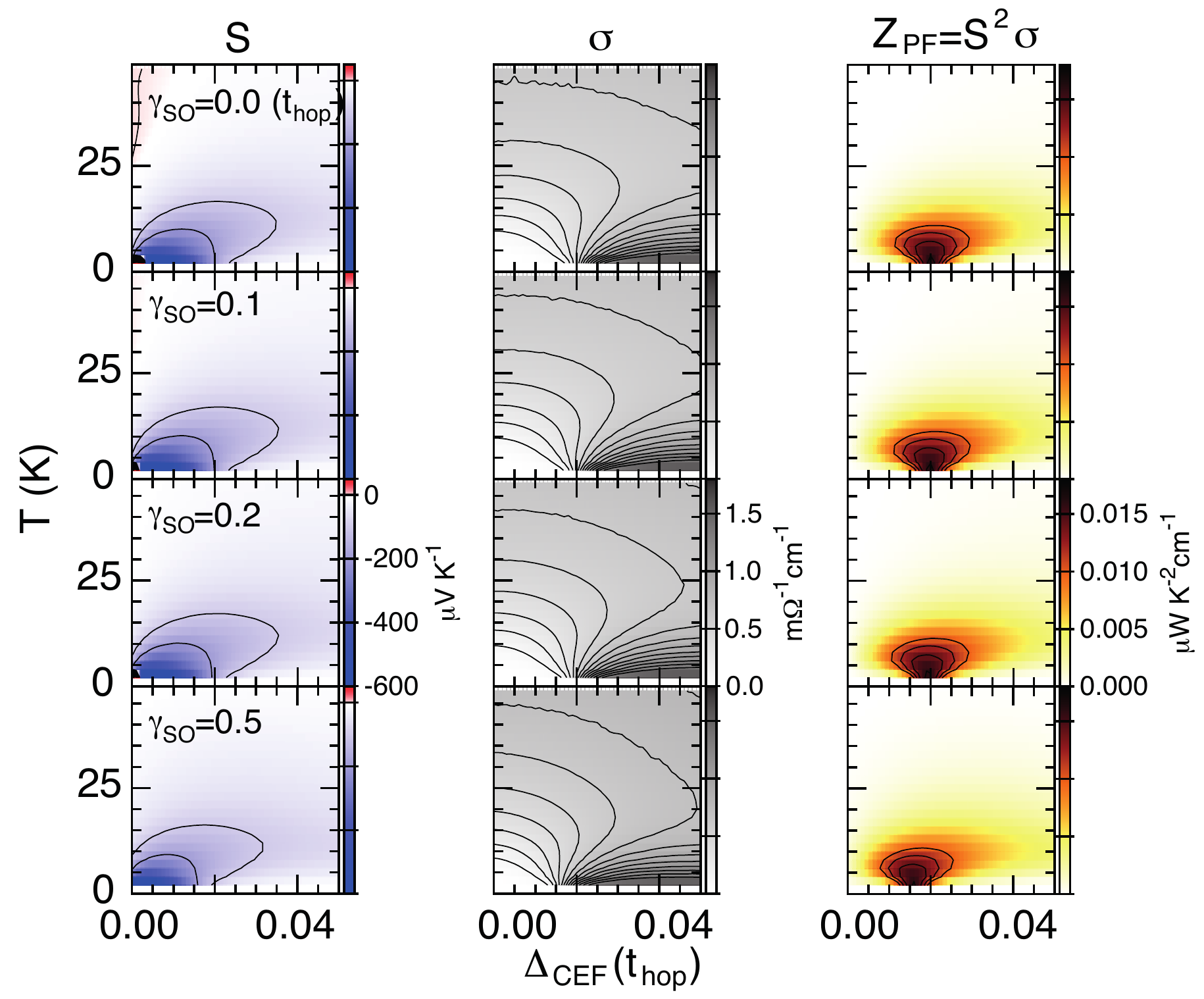}
\caption{Transport properties are compared for each SOI:  $S$ (left column), conductivity $\sig$ (middle), and power factor $Z_{\text{PF}}$ (right). The spin-orbit couplings are chosen to be 0.0, 0.1, 0.2, and 0.5 in units of the hoping amplitude $t_{\text{hop}}$, respectively.}\label{Fig6}
\end{center}
\end{figure}

\begin{figure}[!t]
\begin{center}
\includegraphics[width=0.9\columnwidth]{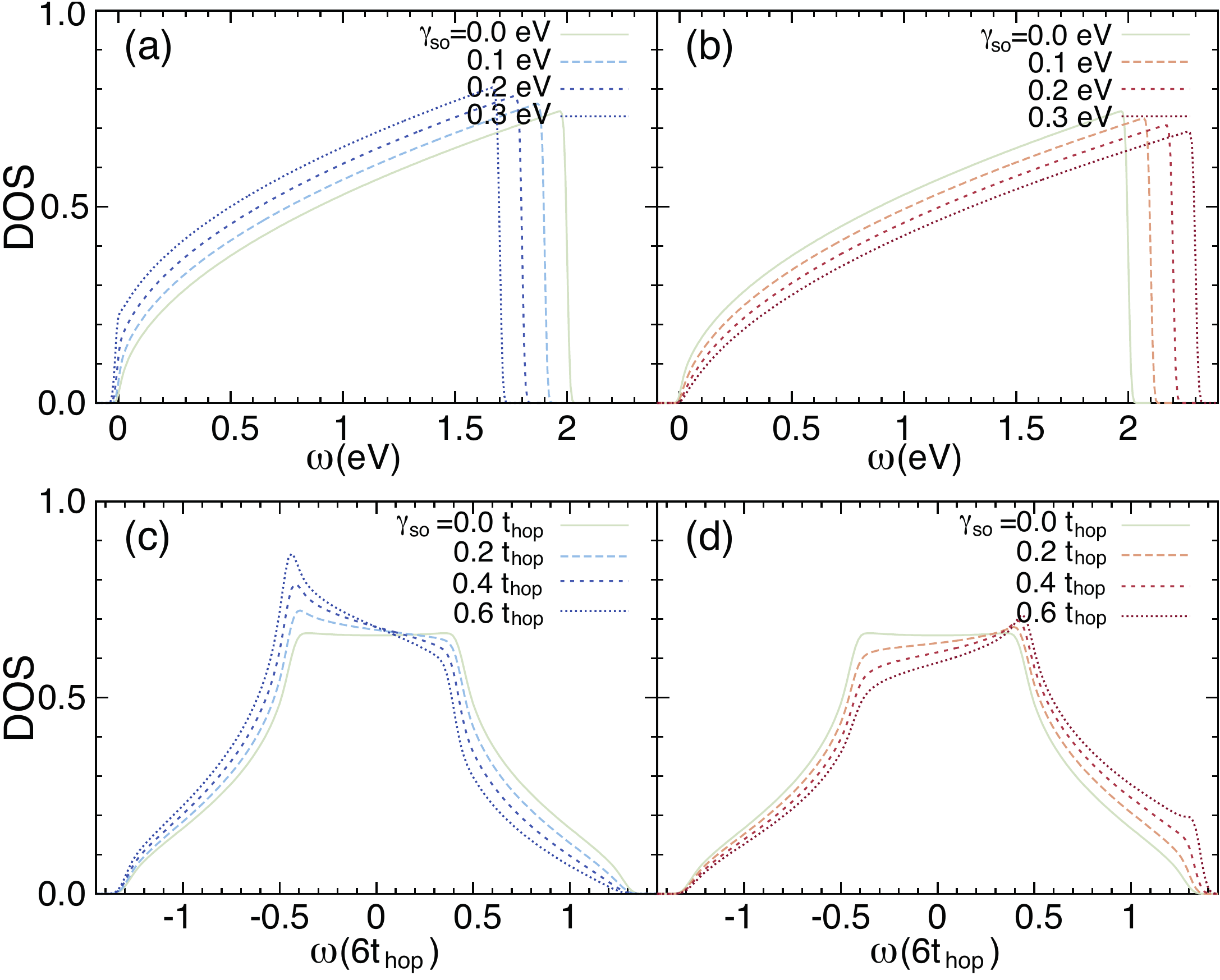}
\caption{Density of states for bare conduction electrons: quadratic
  dispersion(top), 3-D tight binding(bottom), helicity $h=-1$(left,
  blue curves) and $h=1$(right, red lines). For the quadratic
  dispersion, the SOI is taken to be linear in  momentum, as
  $\gamma_{\text{so}}|\mb k|$. The changes in the bandwidth are 
exactly proportional to $\gamma_{\text{so}}$. In figures (c) and (d), the SO term, $\gamma_{\text{so}}\bm\Gamma_{\mb k}$ is taken in accordance with the cubic point group symmetry, and the next nearest neighbor hopping parameter $g_2=0.3$. Note that the bandwidths are not drastically affected by the SO, while the shapes become more asymmetric with the increase in $\gamma_{\text{so}}$.} \label{Fig7}
\end{center}
\end{figure}

From Fig.~\ref{Fig5}, we observe consistency with the parabolic
model: the power factor can be enhanced by adjusting the CEF, while
the SOI slightly lowers the optimal value of the CEF. For $T>15K$, the
enhancement in $Z_{\text{PF}}$ is not as drastic as was in the low
temperature case.  As in the parabolic model, this trend occurs
because the thermally active region is too wide to encompass only one band
[see Fig.\ \ref{Fig4}(c) and (d)]. Comparison with the other
columns reveals that $Z_{\text{PF}}$ is also maximized near an
insulator to a (helical) metal transition (see the conductivity $\sig$
at $T=5K$), which is the consequence of the competition between $S$
and $\sig$. Here, one can observe that $S$ becomes maximal at
$\Del_{\text{CEF}}\simeq 0.01t_{\text{hop}}$, which is a consequence
of the choice $V_2/V_1=1.01$. With $V_2/V_1=1$, $S$ only decreases
with CEF (not shown). Fig.~\ref{Fig6} similarly confirms the
consistency with the parabolic model. The only difference is that the
effect of SOI is not as remarkable, though it works to shift the
optimal value of the CEF. The reason mainly lies in the changes of the
DOS depending on SOI: linearized SOI changes the bandwidth significantly, while 3D tight-binding SOI does not due to its periodic structure. (See Fig.~\ref{Fig7}.)


\subsection{Effect of multi-orbital degeneracy}\label{mr}
In addition to the continuous control of orbital degeneracy through
crystal-field and spin-orbit coupling, we can specifically study the
effect of increasing the number of degenerate orbitals
($l_{\text{max}}=1, 2, \cdots, 5$). To minimize the number of free
parameters, we will set the orbital degeneracy of the two bands (the
allowed values of $l$) to be equal.  Although continuous control is
not possible in this case, one can then consider changing the material
content to achieve a better thermoelectric. Here, the bare conduction
electron dispersion is taken to be that of the 3D tight binding model.

\begin{figure}[!t]
\begin{center}
\includegraphics[width=0.9\columnwidth]{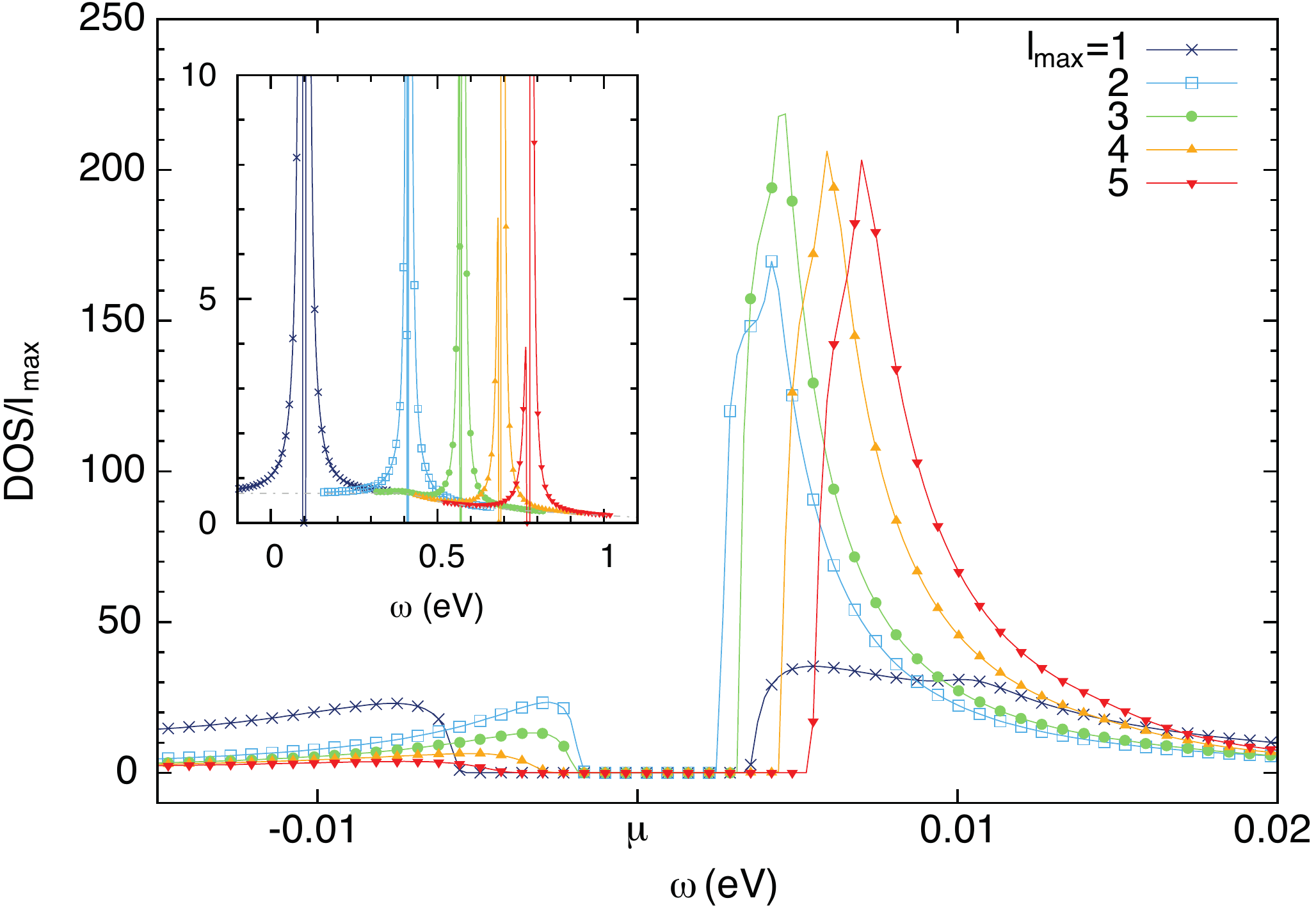}
\caption{Density of states per orbital for different number of orbitals($l_{\text{max}}$), where the chemical potential at each case is adjusted to the center. The inset displays the DOS without the adjustment.} \label{Fig8}
\end{center}
\end{figure}

First, we compare the DOS depending on the number of orbitals involved
[Fig.~\ref{Fig8}]. As $l_{\text{max}}$ increases, the asymmetry
between the upper and the lower bands becomes more pronounced. At the
same time, the insulating gap increases with $l_{\text{max}}$ for
$l_{\text{max}}\geq 2$. Note that this feature is quite similar to the
single impurity problem with $N$ flavors. The inset of Fig.~\ref{Fig8}
displays the DOS without adjusting the chemical potential. Since the
slave-boson method renormalizes the local energy by $\eps_{f}\ra
\eps_{f}+\lambda$, the relative location of the Kondo resonance for
each case (near each gap) indicates that the amount of renormalization
$\lambda$ increases with the number of available orbitals. Since each
band below and above the insulating gap should accommodate one
electron, the deformation of the lower bands becomes less significant
as $l_{\max}$ increases (see the lower bands for different
$l_{\text{max}}$). In other words, since the area below and above the
gap should be equal, the asymmetry of the DOS becomes more significant as $\lambda$, or equivalently $l_{\text{max}}$, increases.

\begin{figure}[!t]
\begin{center}
\includegraphics[width=1.0\columnwidth]{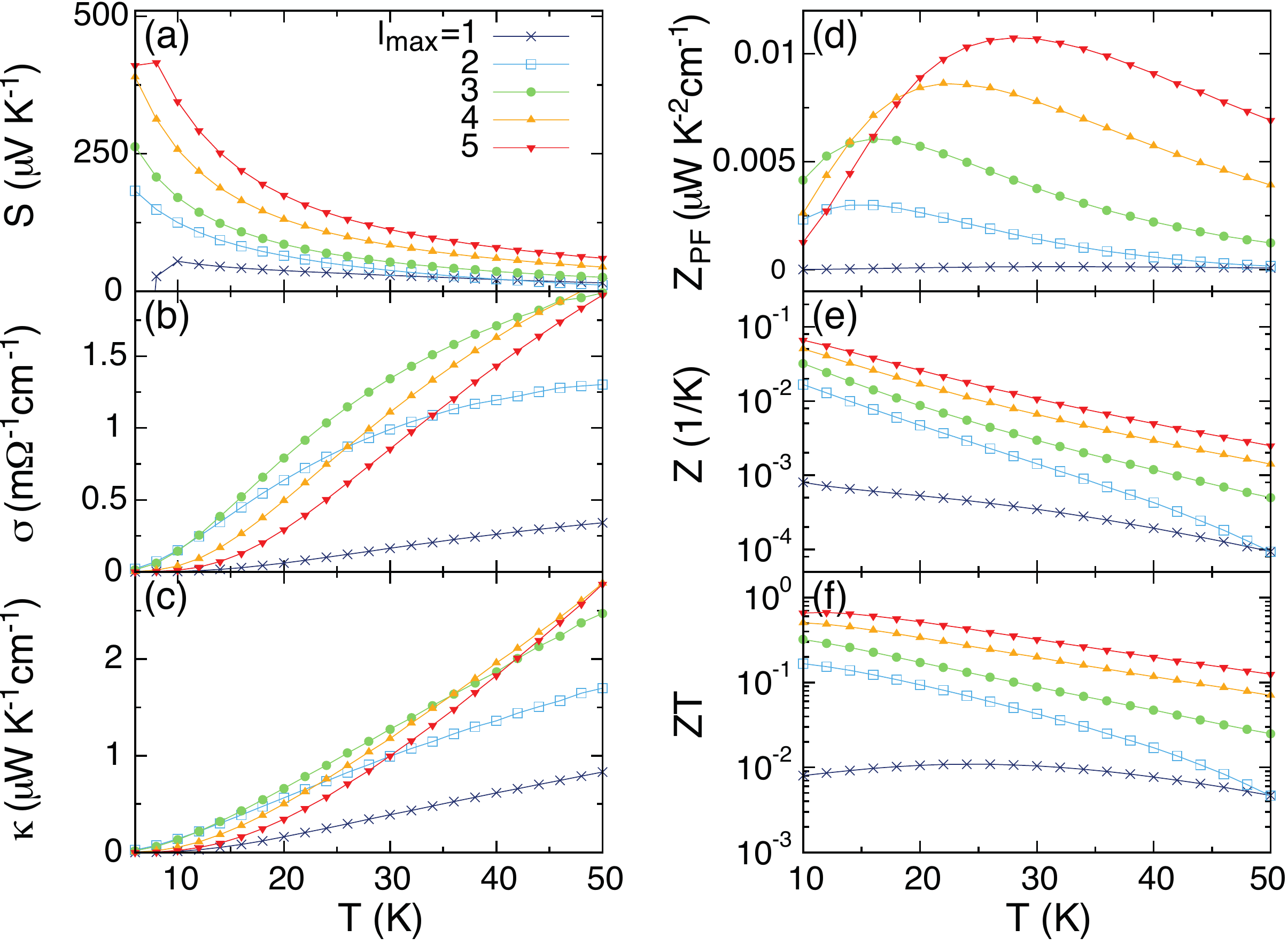}
\caption{Transport properties for $l_{\text{max}}$ orbitals: (a) Seebeck coefficient, (b) electrical conductivity, (c) thermal conductivity, (d) thermopower (e) figure of merit and (f) dimensionless figure of merit. } \label{Fig9}
\end{center}
\end{figure}

Given the band structures at the mean-field level, we proceed to
calculate the transport properties as shown in Fig.~\ref{Fig9}. As
$l_{\text{max}}$ increases, maxima of $Z_{\text{PF}}$ also increases 
as the temperature is elevated. The thermopower is also enhanced
with the number of available orbitals, which is caused by pronounced
asymmetry in the DOS (See Fig.~\ref{Fig8}). Since the gap size
increases, the conductivity generally decreases with the number of
orbitals for $l_{\text{max}}\geq2$. Though not shown here, the maximal
power factor per orbital, $Z_{\text{PF}}/l_{\text{max}}$, also
increases with $l_{\text{max}}$ until $l_{\text{max}}<7$. In
Fig.~\ref{Fig9}(c), the thermal conductivity due to electronic
structure is evaluated, excluding any contribution from lattice
vibrations. Typically, phonons are dominant contributors to the
thermal conductivities, but it may not be so prevalent at the low
temperature range considered here, presumably $T\lesssim 10$K. The
resultant (dimensionless) figure of merit, assuming
$\kappa=\kappa_{\text{electron}}+\kappa_{\text{phonon}}\simeq
\kappa=\kappa_{\text{electron}}$, is strongly enhanced with the number
of orbitals at least by an order of 10.  This is one of the key
results of this paper.

\section{Summary}

We have shown that orbital degeneracy opens a new window into
increasing the power factor for strongly interacting thermoelectrics. Our key findings are
that the power factor is maximized in strongly correlated systems by
tuning  1) the gap between nearly degenerate local f-orbitals through
the crystal field effect, 2) ) the gap between nearly degenerate itinerant electron bands through the spin-orbit coupling,  and 3) the  the number of
degenerate local and itinerant orbitals.  

This approach provides a new parameter space for 
the design of strongly correlated thermoelectric materials. Our result was derived using the slave-particle mean-field theory, which is not expected to be quantitatively reliable but should capture general trends.  The effect of degeneracy in enhancing thermoelectric performance of strongly correlated systems could also be investigated using other methods such dynamical mean-field theory\cite{dmft} and finite frequency methods\cite{shastry}.  Another direction for future work is to consider more profound effects of spin-orbit coupling combined with correlations, as in the proposed ``topological Kondo insulators''~\cite{kondotopological}, whose surface states are currently being sought experimentally; such surface states have the potential to increase thermoelectric performance at low temperatures~\cite{pouyan10}.

Finally, a direction that is difficult theoretically but may be important for actual materials is to find ways of interpolating between the effectively itinerant calculation here (i.e., there are plane-wave states of the slave particles) with the ``atomic limit''~\cite{beni}, where the effects of multiple orbitals have also been considered~\cite{mukerjee}.  The atomic limit, which is valid when the hopping is the smallest energy scale in the problem, has been argued to be relevant to experiments on sodium cobaltates near room temperature~\cite{lee,Mukerjee:2007p23}.  The results of this paper should motivate continued investigation of correlated materials for thermoelectricity and suggest that a guided search with controlled crystal-field splitting may lead to further improvements in thermoelectric figure of merit, especially in the low-temperature regime.


\acknowledgements

This work was supported by the U.S. Department of Energy, Office of
Basic Energy Sciences, Materials Sciences and Engineering Division,
under Contract No. DE-AC02-05CH11231 (P.G. and J.E.M.). P.G. also
acknowledge support from NSF DMR-1064319.  S.H. and P.W.P. are funded by NSF-DMR-1104909.

\appendix

\section{Spin-orbit coupling}\label{sp}
In the presence of SO coupling\cite{SamokhinK09,Isaev}, the conduction electron dispersion matrix $\varepsilon_{\sig\sig'}(\mb k)$ is given by
\beq
\varepsilon_{\sig\sig'}(\mb k) = \eps_{\mb k}\delta_{\sig\sig'}+\bm\Gamma_{\mb k}\cdot\bm\sig_{\sig\sig'},
\eeq
where $\eps_{\mb k}$ is the dispersion without the SO interaction, and
$\bm\sig$ are the Pauli matrices. (The SO interactions considered here
originate from the absence of an inversion symmetry in the crystal
lattice.) The antisymmetric SO coupling is described by the real
pseudovector $\bm\Gamma_{\mb k}$, which is determined by the point
group symmetry of the crystal. For instance, CePt$_3$Si, CeRhSi$_3$
and CeIrSi$_3$ belong to tetragonal point group ($\mbb G=C_{4v}$) in which
\beq\begin{split}
\bm\Gamma_{\mb k} 
	= \gamma_{\text{so}}(& \hat {\mb k}_x\sin{k_ya}
	-\hat {\mb k}_y\sin{k_xa}  \\
	& +\hat {\mb k}_z g_2\sin k_xa\sin k_ya\sin k_zc (\cos k_xa-\cos k_ya ),\\
\end{split}\eeq
in the next nearest neighbor approximation for a real $\gamma_{\text{so}}$, the lattice spacing $a, c$ and the next nearest neighbor parameter $g_2$. In case of the cubic point group symmetry ($\mbb G=O$), the pseudovector is given by
\beq\begin{split}
\bm\Gamma_{\mb k}=\gamma_{\text{so}} & \hat{\mb k}_x\sin k_xa[1-g_2(\cos k_ya+\cos k_za)] \\
	& +(\text{Positive permutations of }x, y, z).
\end{split}\eeq
In real noncentrosymmetric crystals, the typical SO strength ranges up to 200 meV. Instead of working in the spin basis, it is useful to introduce the helical basis that diagonalizes the single-electron dispersion $\varepsilon_{\sig\sig'}(\mb k)\ra \left[U_{\mb k}^\dag\varepsilon(\mb k)U_{\mb k}\right]_{hh'}=\left(\eps_{\mb k}+h|\bm\Gamma_{\mb k}|\right)\delta_{hh'}$ with $h,h'=\pm 1$.
\\

\bibliography{paim}

\begin{thebibliography}{28}%
\makeatletter
\providecommand \@ifxundefined [1]{%
 \@ifx{#1\undefined}
}%
\providecommand \@ifnum [1]{%
 \ifnum #1\expandafter \@firstoftwo
 \else \expandafter \@secondoftwo
 \fi
}%
\providecommand \@ifx [1]{%
 \ifx #1\expandafter \@firstoftwo
 \else \expandafter \@secondoftwo
 \fi
}%
\providecommand \natexlab [1]{#1}%
\providecommand \enquote  [1]{``#1''}%
\providecommand \bibnamefont  [1]{#1}%
\providecommand \bibfnamefont [1]{#1}%
\providecommand \citenamefont [1]{#1}%
\providecommand \href@noop [0]{\@secondoftwo}%
\providecommand \href [0]{\begingroup \@sanitize@url \@href}%
\providecommand \@href[1]{\@@startlink{#1}\@@href}%
\providecommand \@@href[1]{\endgroup#1\@@endlink}%
\providecommand \@sanitize@url [0]{\catcode `\\12\catcode `\$12\catcode
  `\&12\catcode `\#12\catcode `\^12\catcode `\_12\catcode `\%12\relax}%
\providecommand \@@startlink[1]{}%
\providecommand \@@endlink[0]{}%
\providecommand \url  [0]{\begingroup\@sanitize@url \@url }%
\providecommand \@url [1]{\endgroup\@href {#1}{\urlprefix }}%
\providecommand \urlprefix  [0]{URL }%
\providecommand \Eprint [0]{\href }%
\providecommand \doibase [0]{http://dx.doi.org/}%
\providecommand \selectlanguage [0]{\@gobble}%
\providecommand \bibinfo  [0]{\@secondoftwo}%
\providecommand \bibfield  [0]{\@secondoftwo}%
\providecommand \translation [1]{[#1]}%
\providecommand \BibitemOpen [0]{}%
\providecommand \bibitemStop [0]{}%
\providecommand \bibitemNoStop [0]{.\EOS\space}%
\providecommand \EOS [0]{\spacefactor3000\relax}%
\providecommand \BibitemShut  [1]{\csname bibitem#1\endcsname}%
\let\auto@bib@innerbib\@empty
\bibitem [{\citenamefont {Bentien}\ \emph {et~al.}(2007)\citenamefont
  {Bentien}, \citenamefont {Johnsen}, \citenamefont {Madsen}, \citenamefont
  {Iversen},\ and\ \citenamefont {Steglich}}]{bentien07}%
  \BibitemOpen
  \bibfield  {author} {\bibinfo {author} {\bibfnamefont {A.}~\bibnamefont
  {Bentien}}, \bibinfo {author} {\bibfnamefont {S.}~\bibnamefont {Johnsen}},
  \bibinfo {author} {\bibfnamefont {G.~K.~H.}\ \bibnamefont {Madsen}}, \bibinfo
  {author} {\bibfnamefont {B.~B.}\ \bibnamefont {Iversen}}, \ and\ \bibinfo
  {author} {\bibfnamefont {F.}~\bibnamefont {Steglich}},\ }\href@noop {}
  {\bibfield  {journal} {\bibinfo  {journal} {European Physics Letters}\
  }\textbf {\bibinfo {volume} {80}},\ \bibinfo {pages} {17008} (\bibinfo {year}
  {2007})}\BibitemShut {NoStop}%
\bibitem [{\citenamefont {Ghaemi}\ \emph {et~al.}(2010)\citenamefont {Ghaemi},
  \citenamefont {Mong},\ and\ \citenamefont {Moore}}]{pouyan10}%
  \BibitemOpen
  \bibfield  {author} {\bibinfo {author} {\bibfnamefont {P.}~\bibnamefont
  {Ghaemi}}, \bibinfo {author} {\bibfnamefont {R.}~\bibnamefont {Mong}}, \ and\
  \bibinfo {author} {\bibfnamefont {J.~E.}\ \bibnamefont {Moore}},\ }\href@noop
  {} {\bibfield  {journal} {\bibinfo  {journal} {Physical Review Letters}\
  }\textbf {\bibinfo {volume} {105}},\ \bibinfo {pages} {166603} (\bibinfo
  {year} {2010})}\BibitemShut {NoStop}%
\bibitem [{\citenamefont {Poudel}\ \emph {et~al.}(2008)\citenamefont {Poudel}
  \emph {et~al.}}]{dreselhause}%
  \BibitemOpen
  \bibfield  {author} {\bibinfo {author} {\bibfnamefont {B.}~\bibnamefont
  {Poudel}} \emph {et~al.},\ }\href@noop {} {\bibfield  {journal} {\bibinfo
  {journal} {Science}\ }\textbf {\bibinfo {volume} {320}},\ \bibinfo {pages}
  {634} (\bibinfo {year} {2008})}\BibitemShut {NoStop}%
\bibitem [{\citenamefont {Wood}(1988)}]{wood}%
  \BibitemOpen
  \bibfield  {author} {\bibinfo {author} {\bibfnamefont {C.}~\bibnamefont
  {Wood}},\ }\href@noop {} {\bibfield  {journal} {\bibinfo  {journal} {Rep.
  Prog. Phys.}\ }\textbf {\bibinfo {volume} {51}},\ \bibinfo {pages} {459}
  (\bibinfo {year} {1988})}\BibitemShut {NoStop}%
\bibitem [{\citenamefont {Kleinke}(2009)}]{chemrev}%
  \BibitemOpen
  \bibfield  {author} {\bibinfo {author} {\bibfnamefont {H.}~\bibnamefont
  {Kleinke}},\ }\href@noop {} {\bibfield  {journal} {\bibinfo  {journal} {Chem.
  Mater.}\ }\textbf {\bibinfo {volume} {22}},\ \bibinfo {pages} {604} (\bibinfo
  {year} {2009})}\BibitemShut {NoStop}%
\bibitem [{\citenamefont {Anderson}(1961)}]{anderson}%
  \BibitemOpen
  \bibfield  {author} {\bibinfo {author} {\bibfnamefont {P.~W.}\ \bibnamefont
  {Anderson}},\ }\href@noop {} {\bibfield  {journal} {\bibinfo  {journal}
  {Physical Review}\ }\textbf {\bibinfo {volume} {124}},\ \bibinfo {pages} {41}
  (\bibinfo {year} {1961})}\BibitemShut {NoStop}%
\bibitem [{\citenamefont {Hewson}(1993)}]{hewson}%
  \BibitemOpen
  \bibfield  {author} {\bibinfo {author} {\bibfnamefont {A.}~\bibnamefont
  {Hewson}},\ }\href@noop {} {\emph {\bibinfo {title} {The Kondo problem to
  heavy fermions}}}\ (\bibinfo  {publisher} {Cambridge},\ \bibinfo {year}
  {1993})\BibitemShut {NoStop}%
\bibitem [{\citenamefont {Read}\ \emph {et~al.}(1984)\citenamefont {Read},
  \citenamefont {Newns},\ and\ \citenamefont {Doniach}}]{read84}%
  \BibitemOpen
  \bibfield  {author} {\bibinfo {author} {\bibfnamefont {N.}~\bibnamefont
  {Read}}, \bibinfo {author} {\bibfnamefont {D.}~\bibnamefont {Newns}}, \ and\
  \bibinfo {author} {\bibfnamefont {S.}~\bibnamefont {Doniach}},\ }\href@noop
  {} {\bibfield  {journal} {\bibinfo  {journal} {Physical Review B}\ }\textbf
  {\bibinfo {volume} {630}},\ \bibinfo {pages} {3841} (\bibinfo {year}
  {1984})}\BibitemShut {NoStop}%
\bibitem [{\citenamefont {Millis}\ and\ \citenamefont {Lee}(1987)}]{millis87}%
  \BibitemOpen
  \bibfield  {author} {\bibinfo {author} {\bibfnamefont {A.~J.}\ \bibnamefont
  {Millis}}\ and\ \bibinfo {author} {\bibfnamefont {P.~A.}\ \bibnamefont
  {Lee}},\ }\href@noop {} {\bibfield  {journal} {\bibinfo  {journal} {Physical
  Review B}\ }\textbf {\bibinfo {volume} {35}},\ \bibinfo {pages} {3394}
  (\bibinfo {year} {1987})}\BibitemShut {NoStop}%
\bibitem [{\citenamefont {Newns}\ and\ \citenamefont {Read}(1987)}]{read87}%
  \BibitemOpen
  \bibfield  {author} {\bibinfo {author} {\bibfnamefont {D.}~\bibnamefont
  {Newns}}\ and\ \bibinfo {author} {\bibfnamefont {N.}~\bibnamefont {Read}},\
  }\href@noop {} {\bibfield  {journal} {\bibinfo  {journal} {Advances in
  Physics}\ }\textbf {\bibinfo {volume} {36}},\ \bibinfo {pages} {799}
  (\bibinfo {year} {1987})}\BibitemShut {NoStop}%
\bibitem [{\citenamefont {Tomczak}\ \emph {et~al.}(2010)\citenamefont
  {Tomczak}, \citenamefont {Haule}, \citenamefont {Miyake}, \citenamefont
  {Georges},\ and\ \citenamefont {Kotliar}}]{kotliar}%
  \BibitemOpen
  \bibfield  {author} {\bibinfo {author} {\bibfnamefont {J.~M.}\ \bibnamefont
  {Tomczak}}, \bibinfo {author} {\bibfnamefont {K.}~\bibnamefont {Haule}},
  \bibinfo {author} {\bibfnamefont {T.}~\bibnamefont {Miyake}}, \bibinfo
  {author} {\bibfnamefont {A.}~\bibnamefont {Georges}}, \ and\ \bibinfo
  {author} {\bibfnamefont {G.}~\bibnamefont {Kotliar}},\ }\href@noop {}
  {\bibfield  {journal} {\bibinfo  {journal} {Phys. Rev. B}\ }\textbf {\bibinfo
  {volume} {82}},\ \bibinfo {pages} {085104} (\bibinfo {year}
  {2010})}\BibitemShut {NoStop}%
\bibitem [{\citenamefont {Coleman}(2007)}]{piers}%
  \BibitemOpen
  \bibfield  {author} {\bibinfo {author} {\bibfnamefont {P.}~\bibnamefont
  {Coleman}},\ }\enquote {\bibinfo {title} {Heavy fermions: electrons at the
  edge of magnetism},}\ in\ \href@noop {} {\emph {\bibinfo {booktitle}
  {Handbook of Magnetism and Advanced Magnetic Materials.}}},\ \bibinfo
  {editor} {edited by\ \bibinfo {editor} {\bibfnamefont {H.}~\bibnamefont
  {Kronmuller}}\ and\ \bibinfo {editor} {\bibfnamefont {S.}~\bibnamefont
  {Parki}}}\ (\bibinfo  {publisher} {John Wiley and Sons},\ \bibinfo {year}
  {2007})\BibitemShut {NoStop}%
\bibitem [{\citenamefont {Aeppli}\ and\ \citenamefont {Fisk}(1992)}]{kins1}%
  \BibitemOpen
  \bibfield  {author} {\bibinfo {author} {\bibfnamefont {G.}~\bibnamefont
  {Aeppli}}\ and\ \bibinfo {author} {\bibfnamefont {Z.}~\bibnamefont {Fisk}},\
  }\href@noop {} {\bibfield  {journal} {\bibinfo  {journal} {Comm. Condens.
  Matter}\ }\textbf {\bibinfo {volume} {16}},\ \bibinfo {pages} {155} (\bibinfo
  {year} {1992})}\BibitemShut {NoStop}%
\bibitem [{\citenamefont {Riseborough}(2000)}]{kins2}%
  \BibitemOpen
  \bibfield  {author} {\bibinfo {author} {\bibfnamefont {P.}~\bibnamefont
  {Riseborough}},\ }\href@noop {} {\bibfield  {journal} {\bibinfo  {journal}
  {Adv. Phys.r}\ }\textbf {\bibinfo {volume} {49}},\ \bibinfo {pages} {257}
  (\bibinfo {year} {2000})}\BibitemShut {NoStop}%
\bibitem [{\citenamefont {Samokhin}(2009)}]{SamokhinK09}%
  \BibitemOpen
  \bibfield  {author} {\bibinfo {author} {\bibfnamefont {K.~V.}\ \bibnamefont
  {Samokhin}},\ }\href@noop {} {\bibfield  {journal} {\bibinfo  {journal} {Ann.
  Phys.}\ }\textbf {\bibinfo {volume} {324}},\ \bibinfo {pages} {2358}
  (\bibinfo {year} {2009})}\BibitemShut {NoStop}%
\bibitem [{\citenamefont {Isaev}\ \emph {et~al.}(2012)\citenamefont {Isaev},
  \citenamefont {Agterberg},\ and\ \citenamefont {Vekhter}}]{Isaev}%
  \BibitemOpen
  \bibfield  {author} {\bibinfo {author} {\bibfnamefont {L.}~\bibnamefont
  {Isaev}}, \bibinfo {author} {\bibfnamefont {D.~F.}\ \bibnamefont
  {Agterberg}}, \ and\ \bibinfo {author} {\bibfnamefont {I.}~\bibnamefont
  {Vekhter}},\ }\href@noop {} {\bibfield  {journal} {\bibinfo  {journal}
  {Physical Review B}\ }\textbf {\bibinfo {volume} {85}},\ \bibinfo {pages}
  {081107(R)} (\bibinfo {year} {2012})}\BibitemShut {NoStop}%
\bibitem [{\citenamefont {Ghaemi}\ \emph {et~al.}(2008)\citenamefont {Ghaemi},
  \citenamefont {Senthil},\ and\ \citenamefont {Coleman}}]{pouyan08}%
  \BibitemOpen
  \bibfield  {author} {\bibinfo {author} {\bibfnamefont {P.}~\bibnamefont
  {Ghaemi}}, \bibinfo {author} {\bibfnamefont {T.}~\bibnamefont {Senthil}}, \
  and\ \bibinfo {author} {\bibfnamefont {P.}~\bibnamefont {Coleman}},\
  }\href@noop {} {\bibfield  {journal} {\bibinfo  {journal} {Physical Review
  B}\ }\textbf {\bibinfo {volume} {77}},\ \bibinfo {pages} {245108} (\bibinfo
  {year} {2008})}\BibitemShut {NoStop}%
\bibitem [{\citenamefont {Mao}\ and\ \citenamefont {Bedell}(1999)}]{mao99}%
  \BibitemOpen
  \bibfield  {author} {\bibinfo {author} {\bibfnamefont {W.}~\bibnamefont
  {Mao}}\ and\ \bibinfo {author} {\bibfnamefont {K.}~\bibnamefont {Bedell}},\
  }\href@noop {} {\bibfield  {journal} {\bibinfo  {journal} {Physical Review
  B}\ }\textbf {\bibinfo {volume} {59}},\ \bibinfo {pages} {R15590} (\bibinfo
  {year} {1999})}\BibitemShut {NoStop}%
\bibitem [{\citenamefont {Ashcroft}\ and\ \citenamefont {Mermin}(1976)}]{ash}%
  \BibitemOpen
  \bibfield  {author} {\bibinfo {author} {\bibfnamefont {N.~M.}\ \bibnamefont
  {Ashcroft}}\ and\ \bibinfo {author} {\bibfnamefont {N.~D.}\ \bibnamefont
  {Mermin}},\ }\href@noop {} {\emph {\bibinfo {title} {Solid State Physics}}}\
  (\bibinfo {year} {1976})\BibitemShut {NoStop}%
\bibitem [{\citenamefont {Kontani}(2004)}]{kontani}%
  \BibitemOpen
  \bibfield  {author} {\bibinfo {author} {\bibfnamefont {H.}~\bibnamefont
  {Kontani}},\ }\href@noop {} {\bibfield  {journal} {\bibinfo  {journal}
  {Journal of Physical Society of Japan}\ }\textbf {\bibinfo {volume} {73}},\
  \bibinfo {pages} {515} (\bibinfo {year} {2004})}\BibitemShut {NoStop}%
\bibitem [{\citenamefont {Sanchez-Castro}(1996)}]{Carlos96}%
  \BibitemOpen
  \bibfield  {author} {\bibinfo {author} {\bibfnamefont {C.}~\bibnamefont
  {Sanchez-Castro}},\ }\href@noop {} {\bibfield  {journal} {\bibinfo  {journal}
  {Philos. Mag. B}\ }\textbf {\bibinfo {volume} {73}},\ \bibinfo {pages} {525}
  (\bibinfo {year} {1996})}\BibitemShut {NoStop}%
\bibitem [{\citenamefont {Georges}\ and\ \citenamefont {Kotliar}(1992)}]{dmft}%
  \BibitemOpen
  \bibfield  {author} {\bibinfo {author} {\bibfnamefont {A.}~\bibnamefont
  {Georges}}\ and\ \bibinfo {author} {\bibfnamefont {G.}~\bibnamefont
  {Kotliar}},\ }\href@noop {} {\bibfield  {journal} {\bibinfo  {journal}
  {Physical Review B}\ }\textbf {\bibinfo {volume} {45}},\ \bibinfo {pages}
  {6479} (\bibinfo {year} {1992})}\BibitemShut {NoStop}%
\bibitem [{\citenamefont {Shatry}(2009)}]{shastry}%
  \BibitemOpen
  \bibfield  {author} {\bibinfo {author} {\bibfnamefont {B.~S.}\ \bibnamefont
  {Shatry}},\ }\href@noop {} {\bibfield  {journal} {\bibinfo  {journal} {Rep.
  Prog. Phys.}\ }\textbf {\bibinfo {volume} {72}},\ \bibinfo {pages} {016501}
  (\bibinfo {year} {2009})}\BibitemShut {NoStop}%
\bibitem [{\citenamefont {Dzero}\ \emph {et~al.}(2010)\citenamefont {Dzero},
  \citenamefont {Sun}, \citenamefont {Galitski},\ and\ \citenamefont
  {Coleman}}]{kondotopological}%
  \BibitemOpen
  \bibfield  {author} {\bibinfo {author} {\bibfnamefont {M.}~\bibnamefont
  {Dzero}}, \bibinfo {author} {\bibfnamefont {K.}~\bibnamefont {Sun}}, \bibinfo
  {author} {\bibfnamefont {V.}~\bibnamefont {Galitski}}, \ and\ \bibinfo
  {author} {\bibfnamefont {P.}~\bibnamefont {Coleman}},\ }\href@noop {}
  {\bibfield  {journal} {\bibinfo  {journal} {Phys. Rev. Lett.}\ }\textbf
  {\bibinfo {volume} {104}},\ \bibinfo {pages} {106408} (\bibinfo {year}
  {2010})}\BibitemShut {NoStop}%
\bibitem [{\citenamefont {Beni}(1974)}]{beni}%
  \BibitemOpen
  \bibfield  {author} {\bibinfo {author} {\bibfnamefont {G.}~\bibnamefont
  {Beni}},\ }\href@noop {} {\bibfield  {journal} {\bibinfo  {journal} {Phys.
  Rev. B}\ }\textbf {\bibinfo {volume} {10}},\ \bibinfo {pages} {2186}
  (\bibinfo {year} {1974})}\BibitemShut {NoStop}%
\bibitem [{\citenamefont {Mukerjee}(2005)}]{mukerjee}%
  \BibitemOpen
  \bibfield  {author} {\bibinfo {author} {\bibfnamefont {S.}~\bibnamefont
  {Mukerjee}},\ }\href@noop {} {\bibfield  {journal} {\bibinfo  {journal}
  {Phys. Rev. B}\ }\textbf {\bibinfo {volume} {72}},\ \bibinfo {pages} {195109}
  (\bibinfo {year} {2005})}\BibitemShut {NoStop}%
\bibitem [{\citenamefont {Lee}\ \emph {et~al.}(2006)\citenamefont {Lee} \emph
  {et~al.}}]{lee}%
  \BibitemOpen
  \bibfield  {author} {\bibinfo {author} {\bibfnamefont {M.}~\bibnamefont
  {Lee}} \emph {et~al.},\ }\href@noop {} {\bibfield  {journal} {\bibinfo
  {journal} {Nature Materials}\ }\textbf {\bibinfo {volume} {5}},\ \bibinfo
  {pages} {537} (\bibinfo {year} {2006})}\BibitemShut {NoStop}%
\bibitem [{\citenamefont {Mukerjee}\ and\ \citenamefont
  {Moore}(2007)}]{Mukerjee:2007p23}%
  \BibitemOpen
  \bibfield  {author} {\bibinfo {author} {\bibfnamefont {S.}~\bibnamefont
  {Mukerjee}}\ and\ \bibinfo {author} {\bibfnamefont {J.}~\bibnamefont
  {Moore}},\ }\href@noop {} {\bibfield  {journal} {\bibinfo  {journal} {Applied
  Physics Letters}\ }\textbf {\bibinfo {volume} {90}},\ \bibinfo {pages}
  {112107} (\bibinfo {year} {2007})}\BibitemShut {NoStop}%
\end{thebibliography}%

\end{document}